\documentclass[twocolumn,showpacs,preprintnumbers,amsmath,amssymb]{revtex4}

\usepackage{graphicx}
\usepackage{dcolumn}
\usepackage{bm}

\begin{document}


\title{Propagation of Extremely-high Energy Leptons in the Earth:
\\Implications to their detection by the IceCube Neutrino Telescope}

\author{Shigeru Yoshida}
\email{syoshida@hepburn.s.chiba-u.ac.jp}
\homepage{http://www-ppl.s.chiba-u.jp/~syoshida}
\author{Rie Ishibashi}%
\altaffiliation[Now at ]{Ushio Denki, Co.Ltd.}
\author{Hiroko Miyamoto}%
\affiliation{%
Department of Physics, Faculty of Science, Chiba University\\
Chiba 263-8522, Japan
}%

\date{\today}

\begin{abstract}
We present 
the results of numerical calculations on propagation of 
Extremely-high energy (EHE) neutrinos and charged leptons 
in the earth for trajectories in all phase space
of nadir angles.
Our comprehensive calculation
has shown that not only the secondary produced muons
but also taus survive without decaying 
in energy range of 10PeV-100PeV with
intensity approximately three orders of magnitude lower than
the neutrino flux regardless of EHE neutrino production models.
They form detectable horizontal or downgoing events 
in a 1km$^3$ underground neutrino telescope such as the IceCube detector. 
The event rate and the resultant detectability of
EHE signals in comparison with the atmospheric muon
background are also evaluated. 
The 90 \% C.L. upperlimit of
EHE neutrino fluxes by a km$^2$ detection area
would be placed at
 $E^2dF/dE\simeq 3.7\times 10^{-8}$ GeV/cm$^2$ sec sr for $\nu_{\mu}$ 
and $4.6\times 10^{-8}$ for $\nu_{\tau}$ with energies of $10^9$ GeV
in absence of signals with energy-loss in a detection volume of 
10PeV or greater.
\end{abstract}

\maketitle

\section{\label{sec:intro} Introduction}

It is well known that there exist extremely high energy (EHE)
particles in the Universe with energies up to 
$\sim 10^{20}$ eV~\cite{nagano01}.
These EHE cosmic rays (EHECRs) may be originated in and/or producing
neutrinos by the various mechanism. For example collisions of EHECRs
and CMB photons photoproduce cosmogenic neutrinos~\cite{berezinsky69}, 
a consequence from the process known 
as the Greisen-Zatsepin-Kuzmin (GZK) mechanism~\cite{GZK}. 
Possible production of
EHECRs in the present Universe
due to the annihilation or collapse of topological defects (TDs)
such as monopoles and/or cosmic strings~\cite{BHS}
could also generate
EHE neutrinos with energies even reaching 
to the GUT scale~\cite{SLSC,SLBY}.
EHE neutrinos provide, therefore, an unique probe to explore
ultra-high energy Universe, which is one of the center piece of
high energy neutrino astrophysics.

It has been argued that
underground neutrino telescopes being operated and/or planned
to be built are capable of detecting such EHE neutrinos~\cite{halzen01}.
In their travel in the earth to the detection volume
in a telescope, EHE neutrinos collide with nuclei in the rock
due to enhancement of the cross section
at EHE range and produce secondary leptons like muons and taus.
The expected mean free path is 
$\sim 600 (\rho_{rock}/2.65 {\rm g\ cm^{-3}})^{-1} 
(\sigma_{\nu}/10^{-32} {\rm cm^2})^{-1}$ km which is by far shorter
than a typical path length of the propagation in the earth. Moreover,
the decay lifetime is long enough at EHEs for the produced
$\mu$ and $\tau$ to survive and possibly reach the detection volume
directly. Successive reactions of the interactions and decaying
are likely to occur in their propagation and
the propagation processes of EHE particles are rather
complex. The accurate understanding of the EHE neutrino and charged lepton
propagation in the earth is, thus, inevitable for EHE neutrino search
by underground neutrino telescopes. 

There have been considerable discussions in the literature from this point
of view. In the Ref.~\cite{reno03}, the transport equations mainly focusing on
$\nu_{\tau}$ and $\tau$ were solved
and the resultant particle fluxes after the propagation have been
shown for trajectories of several nadir angles in the horizontal
directions such as $85^{\circ}$. It is true that a major faction
of EHE $\tau$ tracks are coming from the horizontal directions
because the earth is opaque for EHE neutrinos, but
a km$^3$ scale neutrino observatory like IceCube is essentially
a 4$\pi$ detector with comparable 
sensitivities to both muons and taus, 
and calculation of EHE particle energy spectra 
of both muons and taus
over all solid angle space including downward event trajectories
would be important to evaluate
detectability with reasonable accuracy. 
Furthermore, they utilized the often-used continuous energy loss (CEL)
approximation that follows only the leading cascade particles.
It is a good approximation for taus, but
the secondary particle fluxes contributed from
the non-leading particles are not negligible for
muons at EHEs where their decay does
not play a visible role. 
Calculations on the earth-skimming EHE $\nu_{\tau}$ have also been made
in some details~\cite{feng02}. 
They used approximations to neglect
contributions of generated leptons from tau interactions/decay
in the earth,
which would be valid enough for consideration of 
earth-skimming neutrino-induced
air showers. It has been pointed out, however, that
the secondary produced $\nu_e$ and $\nu_{\mu}$
by tau decays would also enhance the total neutrino flux~\cite{beacom02},
which would be a benefit for an underground neutrino observatory.
Following all propagating leptons and taking into account
the contributions from particles not only skimming
but propagating deeper in the earth are, therefore, essential
to an underground-based neutrino observatory.

In this work, we numerically calculate
the intensity and the energy distribution of EHE neutrinos and
their secondary produced $\mu$'s and $\tau$'s 
during the propagation in the earth
for the application to a km$^3$ scale neutrino observatory.
The resultant fluxes are shown as a function of nadir angles
from downward to upward going directions. All the relevant interactions
are taken into account and
we follow {\it all} produced particles in the reactions
whereas the CEL approximation follows
only the leading cascade particles.
The initial flux
is mainly assumed to be a bulk of the cosmogenic neutrinos, 
generated from the decay of pions photoproduced by EHE cosmic ray
protons colliding with the cosmic thermal background photons,
since the cosmogenic neutrino model is appropriate for a benchmark as 
the flux prediction is on the solid theoretical foundation.
Its implications to the detection by 
the IceCube neutrino telescope~\cite{IceCube},
which are currently under construction at Antarctica,
are then discussed in some details.

The paper is outlined as follows: First we briefly
review the interactions/decay channels involved with
EHE particle propagation in the earth in Sec.~\ref{sec:interactions}.
The method of our numerical calculations is also briefly explained.
In the Sec.~\ref{sec:results} we show the calculated results:
energy distributions and intensities of muons, taus, and neutrinos
after their propagation. The energy spectra of these EHE particles
are shown for the cosmogenic neutrino model. Implications on the detection
by the IceCube neutrino telescope are discussed in the Sec.~\ref{sec:detection}
and the detectability considering the possible background in
the experiment is discussed in detail. The sensitivity to EHE neutrino
fluxes by a km$^3$ neutrino observatory is also shown.
We summarize our conclusions and make suggestions for future work
in Sec.~\ref{sec:summary}.

\section{\label{sec:interactions} Dynamics of the Propagation in the Earth}

\begin{figure*}\
\includegraphics[width=.8\textwidth,clip=true]{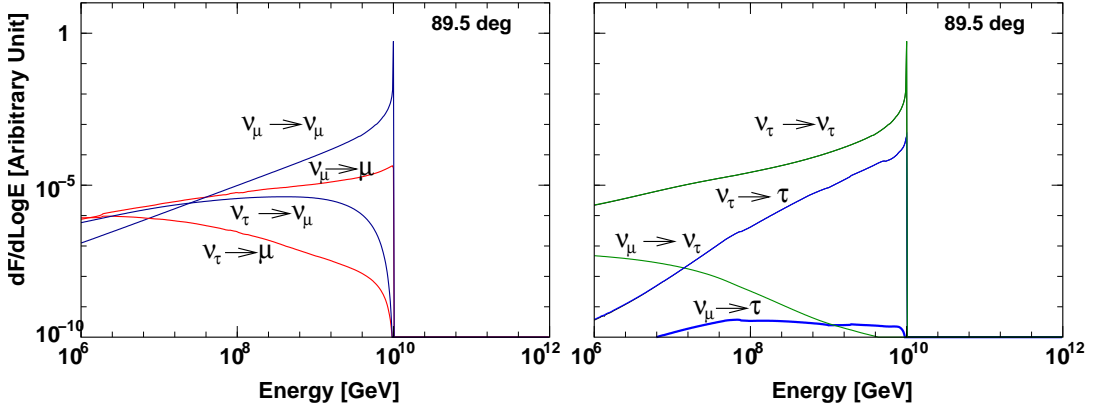}
\caption{\label{fig:propMtx}The energy distribution
of EHE leptons after propagation in the earth with nadir angle
of 89.5 degree. $\mu$'s and $\tau$'s are secondary produced.
The left panel shows the distributions of leptons with $\mu$ flavor
while the right panel shows the case of  $\tau$ flavor.
Input spectrum is $10^{10}$ GeV monochromatic of $\nu_{\mu}$
and $\nu_{\tau}$ with equal intensity of ``1'' in this arbitrary
unit.}
\end{figure*}

EHE neutrinos during the propagation do not penetrate
the earth but are involved in
charged/neutral current interactions
that generate charged leptons and hadronic showers 
because their cross sections are expected to be
enhanced in the ultra-high energy regime. Secondary produced
$\mu$'s and $\tau$'s travel in the earth initiating many radiative reactions
to lose their energy. 
The higher-order interactions like $\mu^\pm$ pair production~\cite{kelner00}
and the charged current disappearance reactions like ${\mu}N\to \nu_{\mu}X$
would regenerate charged leptons and neutrinos which are subject to further 
interactions. Moreover, the $\tau$'s decay channels 
like $\tau\to\nu_{\tau}\mu\nu_{\mu}$ regenerate $\nu_{\tau}$. 
A primary EHE neutrino particle, therefore, results in number of particles
with various energies and species which would be passing through
an instrumented volume of an underground neutrino telescope.
The resultant energy spectra and their intensity 
are consequences from the chain processes of interactions and decay.
Table~\ref{tab:Chain} summarizes the interaction/decay channels
as a function of primary and generated particle species.
Main energy loss process for secondary produced $\mu$'s and $\tau$'s
are, $e^\pm$ pair creation, Bremsstrahlung, and 
the photonuclear interactions.
The relevant cross sections are formulated in 
Ref.~\cite{PC} for pair creation, 
Ref.~\cite{BS} for Bremsstrahlung,
Ref.~\cite{PN} for photonuclear interaction. 
Among them the photonuclear cross section
has the largest theoretical uncertainty because it relies on the details
of the nuclei structure function, which has to be estimated
from extrapolation from the low energy data.
In the present calculation is used the estimation based on 
the deep-inelastic scattering formalism with the ALLM parameterization 
of the structure function~\cite{ALLM}, 
which has been considered to be most reliable
prediction.  We artificially switch off the photonuclear interaction
to see its systematic uncertainty in the results later in this paper.
Furthermore, the week interaction, $l^{\pm}N\to \nu X$,
to cause muon and tau disappearances, and the heavier lepton pair production
such as $\mu^+\mu^-$~\cite{kelner00} 
are also taken into account in the present calculation,
which leads to a visible contribution to the particle fluxes at EHEs.

An EHE neutrino is a subject to charged-current (CC) and neutral-current (NC)
interactions with nucleon.
As there is no direct measurement of the relevant interactions in EHE range,
the predictions of the $\nu N$ cross sections rely on
incompletely tested assumptions about the behavior of parton distributions
at very small values of the momentum fraction $x$.
Since we do not have further clues to investigate EHE neutrino interactions
in our hands, we limit our present analysis within the range of the standard 
particle physics and use the cross section estimated by
Ref.~\cite{gandhi96} using
the CTEQ version 5 parton distribution functions~\cite{cteq}.

\begin{table}
\caption{\label{tab:Chain} Interactions and decay channels
involved in the EHE particle propagation in the earth.
Rows are primary and columns are generated particles.}
\begin{ruledtabular}
\begin{tabular}{c|ccccccc}
      & $\nu_e$ & $\nu_\mu$ & $\nu_\tau$ & e/$\gamma$
      & $\mu$ & $\tau$ & hadron \\\hline
$\nu_e$ & NC\footnotemark[1] & & & CC\footnotemark[2] & & & CC/NC \\
$\nu_\mu$ & & NC & & & CC & & CC/NC\\
$\nu_\tau$ & & & NC & & & CC & CC/NC\\
$\mu$ & D\footnotemark[3] & D/CC & & 
 P\footnotemark[4]/B\footnotemark[5]/D & P & P& PN\footnotemark[6]/CC\\ 
$\tau$ & D & D & D/CC & P/B/D & P/D & P & PN/CC/D \\
\end{tabular}
\end{ruledtabular}
\footnotetext[1]{Neutral Current interaction.}
\footnotetext[2]{Charged Current interaction.}
\footnotetext[3]{Decay.}
\footnotetext[4]{Pair Creation.}
\footnotetext[5]{Bremsstrahlung.}
\footnotetext[6]{Photonuclear interaction.}
\end{table}

Decay processes are also major channels
and compete with the interaction processes depending on energy.
The $\mu$ and $\tau$-leptonic decay distribution can be
analytically calculated from the decay matrix
using the approximation that the generated lepton mass
is negligible compared to that of the parent lepton~\cite{gaisser90}.
For $z=E_{\nu_l}/E_l$ ($l=\mu$ or $\tau)$ it is written
by

\begin{equation}
{dn\over dz} = {5\over 3}-3z^2+{4\over 3}z^3
-\left({1\over 3}-3z^2+{8\over 3}z^3\right),
\label{eq:decay}
\end{equation}

and for $y_\nu = E_{\nu_e}/E_\mu$ ($\mu$ decay), $E_{\nu_{e,\mu}}/E_\tau$
($\tau$ decay), 

\begin{equation}
{dn\over dz} = 2-6y_{\nu}^2+4y_{\nu}^3 
-\left(-2+12y_{\nu}-18y_{\nu}^2+8y_{\nu}^3\right).
\label{eq:decay2}
\end{equation}

The hadronic $\tau$ decay has various mode and its accurate
treatment is rather difficult. Here we use the 2-body
decay approximation as in Ref.~\cite{TAU}.

The transport equations to describe the particle propagation
in the earth are given by
{\small
\begin{eqnarray}
{dJ_{\nu}\over dX} &=& -N_A\sigma_{\nu N, CC+NC}J_{\nu}+\
{m_l\over c\rho\tau^d_l}\int dE_l {1\over E_l} 
{dn^d_l\over dE_{\nu}}J_{l}(E_l)\nonumber\\
&& + N_A \int dE^{'}_{\nu} \
{d\sigma_{\nu N, NC}\over dE_{\nu}}J_{\nu}(E^{'}_{\nu})\nonumber\\
&& + N_A \int dE^{'}_{l} \
{d\sigma_{l N, CC}\over dE_{\nu}}J_l(E^{'}_{l})\label{eq:transport1}\\
{dJ_{l}\over dX} &=& -N_A\sigma_{l N}J_{l}-\
{m_l\over c\rho\tau^d_l E_l}J_{l}\nonumber\\
&&+ N_A \int dE^{'}_{\nu} {d\sigma_{\nu N, CC}\over dE_{l}}\
J_{\nu}(E^{'}_{\nu})\nonumber\\
&& + N_A \int dE^{'}_{l} {d\sigma_{l N}\over dE_{l}}J_l(E^{'}_{l})\nonumber\\
&&+{m_l\over c\rho\tau^d_l}\int dE^{'}_l \
{1\over E^{'}_l} {dn^d_l\over dE_l}J_{l}(E^{'}_{l}),
\label{eq:transport2}
\end{eqnarray}
}
where $J_{l}=dN_l/dE_l$ and $J_{\nu}=dN_{\nu}/dE_{\nu}$ are
differential
fluxes of charged leptons and neutrinos, respectively, 
$N_A$ is the Avogadro's number,
$\rho$ is the local density of the medium (rock/ice) 
in the propagation path, 
$\sigma$ is the relevant interaction cross sections, $dn^d_l/dE$
is the energy distribution of the decay products which is derived from
the decay rate per unit energy and given by Eqs.~\ref{eq:decay} 
and~\ref{eq:decay2}, $c$ is the speed of light,
$m_l$ and $\tau^d_l$ are mass and the decay life time of the lepton $l$,
respectively. The density profile of the rock, $\rho(L)$,
is given by the Preliminary Earth Model~\cite{PEM}.
A column density $X$ is defined by $X=\int\limits^L_0 \rho(L^{'})dL^{'}$.

Eq.~\ref{eq:transport1} describes the neutrino propagation.
The first term
is a loss due to the neutrino interaction, the second represents a
contribution due to the decay, the rest of the terms accounts
generation of neutrinos by the neutrino and charged lepton interactions.
The fourth term represents the neutrino appearance by
the CC interactions such as $\mu$N$\to\nu_{\mu}$X.
Eq.~\ref{eq:transport2} describes the charged lepton propagation
and has the similar terms with those of Eq.~\ref{eq:transport1},
but also the term to represent a loss due to the lepton decay.

\begin{figure}\
\includegraphics[width=.4\textwidth,clip=true]{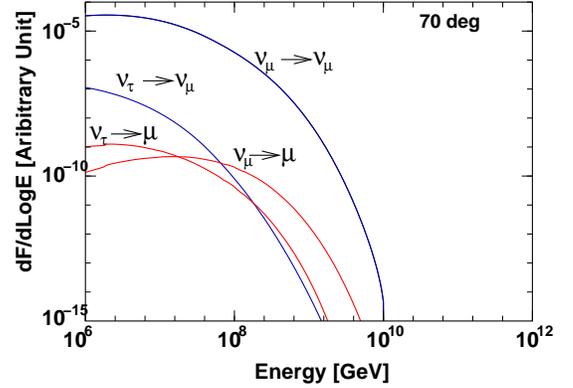}
\caption{\label{fig:propMtx2} Same as Fig.~\ref{fig:propMtx},
but for nadir angle of 70 degree.
The distributions of leptons with $\mu$ flavor
when the input $\nu_\mu$ and $\nu_{\tau}$ is
monochromatic energy of $10^{10}$ GeV are shown.}
\end{figure}

We numerically calculated these equations by building the matrices
describing the particle propagation over infinitesimal distance as
described in Ref~\cite{yoshida93,proth96}. 
The energy differential cross sections are derived from
the ones for the inelasticity parameter $y=1-E^{'}/E$, {\it i.e.},
$d\sigma/dy$.
Let us show two examples to show the behavior of the EHE particle
propagation in the earth. Fig.~\ref{fig:propMtx} shows the energy
distribution of EHE leptons after propagation in the earth
entering with nadir angle of $89.5^{\circ}$. The corresponding
propagation distance in the earth is $\sim 110$ km.
Primary input spectrum is monochromatic 
energy distribution of $10^{10}$ GeV of
$\nu_{\mu}$ and $\nu_{\tau}$ with equal intensity.
Sizable bulks of the secondary produced $\mu$'s and $\tau$'s
are found. As the $\mu$ bulk from $\nu_{\tau}$ are mainly generated
from $\tau$ decay which occurs less frequently in high energy
region, their intensity decreases with higher energy. For the same
reason, the secondary $\tau$'s energy distribution is
harder than that of $\mu$'s. Note that $\tau$ originated in
primary $\nu_{\mu}$ denoted as $\nu_{\mu}\to\tau$ in the right
panel in the figure are produced in heavy lepton pair creation
$\mu\to\mu\tau^{+}\tau^{-}$. 

The intensities of ``prompt'' muon and tau, 
whose energies are approximately same
with the primary neutrinos, are four to five order
of magnitude lower than the primary neutrino flux
as indicated in the figures, but low energy bulk
of the secondary muon and tau which has suffered
energy loss during their propagation makes significant
contribution to the flux for a given neutrino 
energy spectrum. It should also be remarked that
the muons generated from secondary produced tau decay
denoted as $\nu_{\tau}\to\mu$
constitute a major fraction of the intensity
below $10^8$ GeV.
We see in the next section
that they form a sizable flux for the EHE neutrino 
model producing hard energy spectrum like the cosmogenic
neutrinos generated by the GZK mechanism.

When particles are propagating more vertically upward going,
{\it i.e.} their propagation distance is longer,
all the prompt component disappears and no particles
essentially survive in EHE range, because of the
significant energy losses.  A typical case is shown
in Fig.~\ref{fig:propMtx2} for nadir angle of 70 degree.
One can see that most bulk of the secondary muons and neutrinos
are absorbed and remain only in low energy range.

Energy distributions and their intensities
of EHE particles propagating in earth are, consequently,
strongly dependent on the zenith (or nadir) angle
of the trajectory, and also on the initial neutrino
energy spectrum. One must solve the transport equation
in the entire phase space in the zenith angle
in order to make accurate estimations of
fluxes we see in an underground neutrino telescope
for a given neutrino initial flux. 

\section{\label{sec:results} The Cosmogenic Neutrino Flux
at Underground Depth}

\begin{figure}\
\includegraphics[width=.4\textwidth,clip=true]{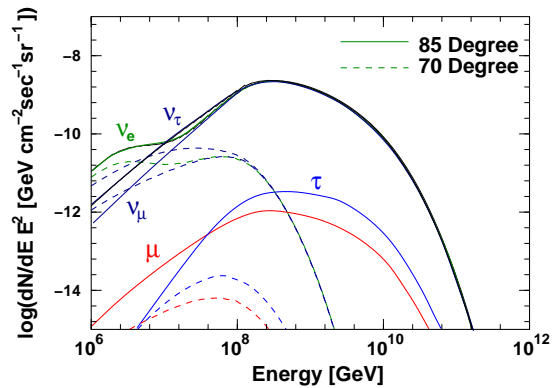}
\caption{\label{fig:gzk_4_4_85Deg}
Fluxes of the EHE particles at the IceCube depth
for a scenario of the neutrino production by the GZK mechanism.
Two cases in the nadir angle are shown in the figure.}
\end{figure}

In this section,
we discuss the case when the initial fluxes of 
$\nu_{\mu}$ and $\nu_{\tau}$ are given by the GZK mechanism, 
EHE neutrino production by collisions of EHECRs to CMB photons
in extragalactic space, as
this model has been thought to be most conventional
mechanism to generate EHE neutrinos without
new physics and/or speculative assumptions.
The biggest uncertainty in the intensity of the cosmogenic
fluxes is related to the cosmic ray source distributions.
Assuming the homogeneously distributed astrophysical sources, however,
variations of the magnitude of the neutrino flux above $10^{9}$ GeV
are restricted approximately within a factor of 10~\cite{seckel01}. 
Although assuming extremely hard cosmic ray injection spectrum 
like $\sim$E$^{-1}$
or very strong source evolution allows larger fluxes 
which can still be consistent with the EHECR and 
the EGRET $\gamma$-ray observations~\cite{kalashev02},
here we limit the present calculations to the conventional case that
the homogeneously distributed astrophysical source
are responsible for the observed EHECR flux below $10^{20}$ eV.

We solve Eqs.~\ref{eq:transport1} and \ref{eq:transport2}
to evaluate the particle fluxes at an underground depth
where a kilometer-scale neutrino observatory is expected to be located.
The IceCube neutrino telescope is constructed at 1400 m depth
below the ice surface and we take this number as a representative
depth. It has been found that changing this depth within a factor
of two would not affect the overall EHE particle intensity 
in significant manner and the conclusion remains same.
The neutrino oscillation with full mixing is assumed 
and $\nu_{\mu}$ initial flux is identical to that
of $\nu_{\tau}$. For the parameters constrained by
the SuperK experiment~\cite{SK00}, the oscillation probability
in the earth in EHE range is negligible, however,
and we do not
account the oscillation in the present calculation on
the propagation.

Fig.~\ref{fig:gzk_4_4_85Deg} shows the fluxes
with nadir angle of 85 and 70 degrees, respectively.
The initial primary cosmogenic neutrino fluxes
are taken from Ref.~\cite{yoshida93}.
Taus notably dominate muons because
their heavy mass makes them penetrate the earth
and because the decay is less important
than interactions for the relevant energy range.
The case of nadir angle of $70^{\circ}$ exhibits the strong
attenuation, however, due to the fact that
mean free paths of all the relevant interactions
including the weak interactions of neutrinos
are by far shorter than the propagation distance.
This implies that most of the up-going events
in a neutrino observatory are coming from horizontal
directions.

\begin{figure}\
\includegraphics[width=.4\textwidth,clip=true]{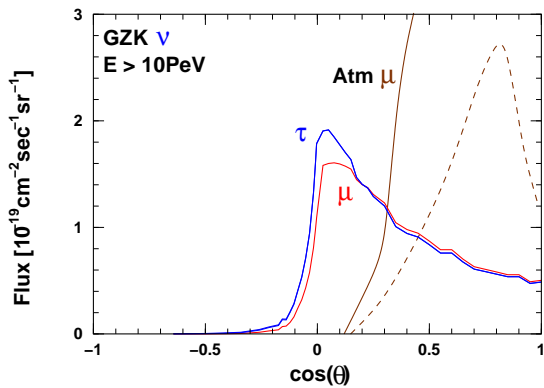}
\caption{\label{fig:atmMuon_angle_10PeV}
Dependence of the muon and tau fluxes originated
in the cosmogenic neutrinos on the cosine of zenith angle.
The integral flux above 10 PeV is plotted in linear scale.
The atmospheric muon fluxes are also shown by 
the solid curve for the conservative estimation
with the low energy extrapolation and by the dashed curve
for the Corsika-based estimation.
The detail of the atmospheric fluxes is discussed in
Sec.~\ref{sec:detection}.}
\end{figure}

$\nu_{\tau}$ flux becomes dominating
over that of $\nu_{\mu}$ in low energy range
where the $\tau$ decay is significantly
more important than interactions.
This enhancement is made by $\nu_{\tau}\to\tau\to\nu_{\tau}$
regeneration process. Note that the small bump
of the $\nu_e$ spectrum is not the propagation
effect but generated primarily by EHECR neutron
decay in the space~\cite{yoshida93,seckel01}.

\begin{figure*}
\includegraphics[width=.8\textwidth,clip=true]{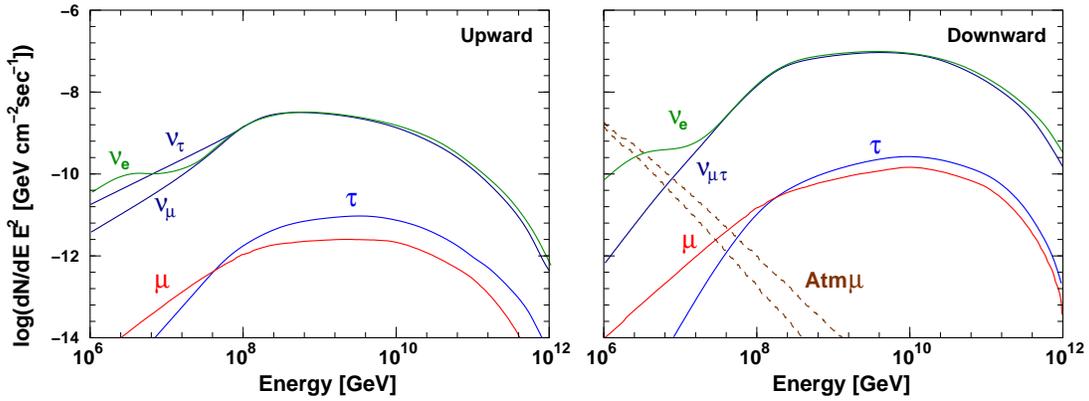}
\caption{\label{fig:gzk_4_4_all}
Energy spectra of $\nu_e,\nu_\mu,\nu_\tau,\mu,\tau$
originated in the cosmogenic neutrinos at
the IceCube depth. The intensities are integrated
over solid angle and shown for the upward region
($\cos\theta\leq 0$: Left panel) and the downward
region ($\cos\theta\geq 0$: Right panel).
The two dashed lines represent the atmospheric muon intensities.
The upper line shows the conservative estimation based on
simple extrapolation from the calculation at 5TeV while
the lower line is derived by the Monte Carlo simulation
with the Corsika package.}
\end{figure*}

The intensity strongly depends on the nadir angle.
Fig.~\ref{fig:atmMuon_angle_10PeV} shows dependences 
of the secondary muon and tau fluxes on the zenith angles.
Strong attenuation by the earth can be seen but
the fluxes are more or less stable in 
the region of the ``downward'' events
where $\cos\theta\geq 0$. Particles in this range
are propagating in ice ($\rho =0.917$ g/cm$^3$) to enter into 
the detection volume. We numerically solved
the transport equation in the ice medium to
derive the downward fluxes. The downward fluxes
constitute major fraction of events in
an underground neutrino observatory. The detection
issues are discussed in Sec.~\ref{sec:detection}.

The energy spectra integrated over zenith angle
are shown in Fig.~\ref{fig:gzk_4_4_all}. 
Secondary muons and taus form a potentially
detectable bulk with intensity of $\sim$ three
order of magnitude lower than the neutrino fluxes.
Main energy range is 10 PeV to 10EeV ($=10^{10}$ GeV).
Regardless of the neutrino production model,
the relative intensity of $\mu$ and $\tau$
to $\nu_{\mu}$ and $\nu_{\tau}$ remains approximately
unchanged. It should be remarked that
intensity of the downward going muons and taus are
larger than the upward ones by an order of magnitude.
As also seen in Fig.~\ref{fig:gzk_4_4_85Deg},
the tau flux dominates over the muons above $10^8$ GeV.
Enhancement of $\nu_{\tau}$ intensity by the regeneration
also appears in the upward going trajectories.

\begin{figure}\
\includegraphics[width=.4\textwidth,clip=true]{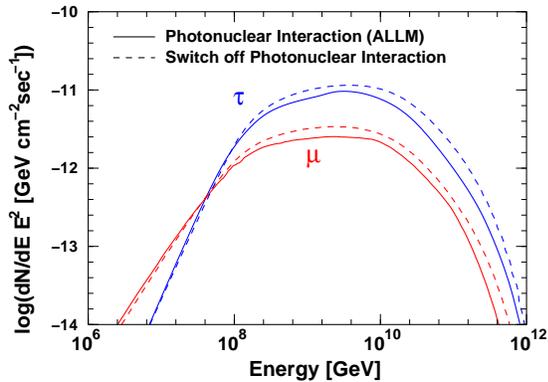}
\caption{\label{fig:gzk_4_4_Lepton}
Dependence of the muon and tau upward fluxes 
on the photonuclear interactions. The integrated
fluxes over nadir angle of $0^{\circ}$ to $90^{\circ}$ are shown.}
\end{figure}

The uncertainty in the muon and tau fluxes estimations
mainly arises from the fact that we do not know
the photonuclear cross section accurately in EHE range.
For example, using
the updated approach to deduce the photonuclear cross section
including a soft part of the photonuclear interaction
would lead to $\sim 30$ \%
enhancement of the total tau energy loss in EHE range\cite{igor2003}.
To be conservative, in Fig.~\ref{fig:gzk_4_4_Lepton} 
we show  the comparison
of the fluxes with and without photonuclear reactions.
Switching off the photonuclear interactions
results in a factor of two variance in the intensity,
which would represent the error range of the secondary tau flux
estimations in a conservative manner.

\section{\label{sec:detection} Detection by 
a kilometer-scale Neutrino Observatory}

\begin{figure}
\includegraphics[width=.4\textwidth,clip=true]{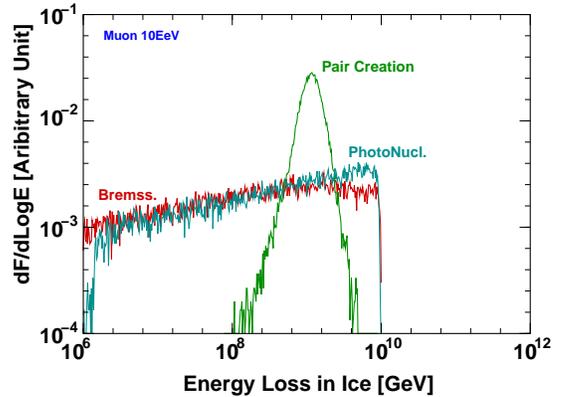}
\caption{\label{fig:mu1km}
Distribution of energy-loss in propagation of muons
over 1km in ice. The primary energy of muons is $10^{10}$ GeV.
Contributions from each interaction are shown separately.}
\end{figure}

The event rate for a neutrino observatory
can be estimated by integrating of the energy spectra
shown in Fig.~\ref{fig:gzk_4_4_all} above a threshold
energy multiplied by an effective area of the detector
which is 1 km$^2$ in case of the IceCube. The downward events
are major contributions and it is necessary to consider
the atmospheric muon background, however.
The atmospheric muon flux estimation in the relevant energy range
is not straightforward because there has been no measurement
available and the numerical calculation is also time-consuming
as one must fully simulate EHE air shower cascades.
Here we use two methods to estimate the flux.
One is to extrapolate the calculations in 5 TeV~\cite{ATM96} which has
been confirmed to be consistent with the measurement.
Because the cosmic ray energy spectrum
follows $E^{-2.7}$ in TeV region while high energy cosmic ray
spectrum above 10 PeV are steeper following $E^{-3}$,
this extrapolation would be overestimating the flux,
but it gives the conservative evaluation.
Another is to run the Corsika air shower simulation~\cite{Corsika}
with energy spectrum of the observed $E^{-3}$ under assumption
that all the mass composition is proton,
and count number of high energy muons reaching to ground.
Then we solve the transport equations for the derived
muon fluxes at surface. The obtained results of the background
intensity in downward events are shown in the right
panel of Fig.~\ref{fig:gzk_4_4_all} by two dashed lines. 
One can see
that the muon background spectrum is quite steep.
Setting a higher threshold energy, therefore,
would be able to eliminate the background contamination.
The flux dependence on the zenith angle is shown
in Fig.~\ref{fig:atmMuon_angle_10PeV} when the threshold
energy is 10 PeV. It is clearly seen that
the muon background attenuates
faster than the neutrino-induced EHE muons and taus,
and there is a window where the signals dominate the muon
background. Table~\ref{tab:event} summarizes
the intensity with threshold energy of 10 PeV
for the various EHE neutrino models together with those
of the atmospheric muon background.

\begin{table*}
\caption{\label{tab:event} Integral flux intensities
for several EHE neutrino models.}
\begin{ruledtabular}
\begin{tabular}{l|ccccc}
   & $I_{\nu_{\mu,\tau}}(E\geq 10PeV)$\footnotemark[1] & 
   $I_{\mu}(E\geq 10PeV)$ & $I_{\tau}(E\geq 10PeV)$ & 
   $I_{\mu}(E_{loss}\geq 10PeV)$ & $I_{\tau}(E_{loss}\geq 10PeV)$\\
   & [cm$^{-2}$ sec$^{-1}$ 2$\pi^{-1}$] & [cm$^{-2}$ sec$^{-1}$] &
   [cm$^{-2}$ sec$^{-1}$] & [cm$^{-2}$ sec$^{-1}$] & [cm$^{-2}$ sec$^{-1}$]\\\hline 
GZK\footnotemark[2] Downward& $5.97\times 10^{-16}$ & 
$5.90\times 10^{-19}$ & $5.97\times 10^{-19}$ & 
$4.75\times 10^{-19}$ & $3.28\times 10^{-19}$ \\
GZK Upward & $5.97\times 10^{-16}$ &  
$3.91\times 10^{-20}$ & $6.63\times 10^{-20}$ & 
$2.57\times 10^{-20}$ & $2.64\times 10^{-20}$ \\
TD\footnotemark[3] Downward & $9.92\times 10^{-15}$ & 
$5.48\times 10^{-18}$ & $5.11\times 10^{-18}$ & 
$3.75\times 10^{-18}$ & $2.94\times 10^{-18}$\\
Atmospheric $\mu$ & - & 
$2.06\times 10^{-18}$ & - & 
$1.74\times 10^{-19}$ & - \\
Atmospheric $\mu$\footnotemark[4] & - & 
$7.25\times 10^{-19}$ & - & 
$5.34\times 10^{-20}$ & - \\
\end{tabular}
\end{ruledtabular}
\footnotetext[1]{Intensity at surface before propagating in the earth.}
\footnotetext[2]{Cosmogenic neutrinos with 
(m,Z$_{max}$) = (4.0,4.0) in Ref.~\cite{yoshida93}.}
\footnotetext[3]{Topological Defects scenario
using SUSY-based fragmentation function in Ref.~\cite{SLBY}}
\footnotetext[4]{Estimation based on the Corsika simulation}
\end{table*}

\begin{figure}
\includegraphics[width=.4\textwidth,clip=true]{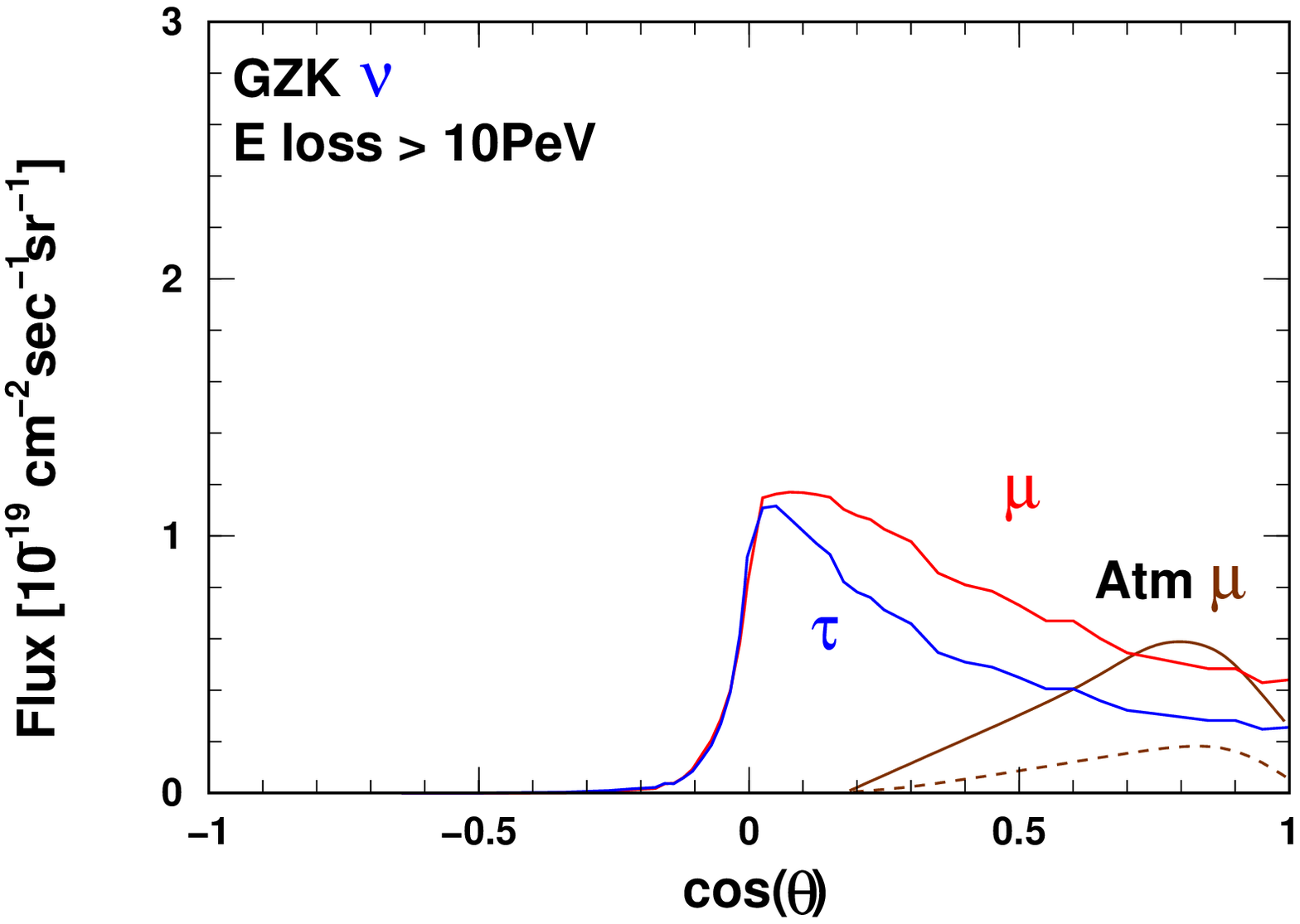}
\caption{\label{fig:atmMuon_casMu_angle_10PeV}
Dependence of the muon and tau fluxes originated
in the cosmogenic neutrinos on the cosine of zenith angle.
The integral flux above 10 PeV of the energy loss 
is plotted in linear scale.
The atmospheric muon fluxes are also shown by 
the solid curve for the conservative estimation
with the low energy extrapolation and by the dashed curve
for the Corsika-based estimation.}
\end{figure}

In fact, what neutrino detectors can measure in direct manner
is not energy of muon/tau tracks but energy loss
in a detection volume. The relation between energy 
and energy loss is approximately $-dE/dX \sim \beta E$.
Here $\beta$ is the average inelasticity
given by
\begin{equation}
\beta = \int\limits_{y_{min}}^{y_{max}}dy^{'}y^{'}{d\sigma\over dy^{'}}.
\label{eq:inelasticity}
\end{equation}
For the $e^{\pm}$ pair creation of muons in ice, 
$\beta\simeq 1.3\times 10^{-6}$ cm$^2$/g.
Therefore the average energy loss fraction due to the pair creation
is
\begin{eqnarray}
{\Delta E\over E} & \simeq & \beta^{e^\pm} \Delta X\nonumber\\
 =  & 0.12 &\
\left({\beta^{e^\pm}\over 1.3\times 10^{-6}}\right)\
\left({\rho_{ice}\over 0.92 {\rm g} {\rm cm}^{-2}}\right)\
\left({\Delta L\over 1 {\rm km}}\right)
\label{eq:eloss_pair}
\end{eqnarray}
indicating that 10 \% of the muon primary energy is deposited
in a detection volume. 
Because the radiative interactions like Bremsstrahlung
have stochastic nature, $\Delta$ E is fluctuated significantly
in event by event bases, however.
We carried out a Monte Carlo simulation to see
the fluctuation. The simulation code uses the same
cross section and the decay tables but calculates an energy
of a particle after an infinitesimal propagation length
$\Delta X$ with the Monte Carlo method instead of solving
the transport equations.  Fig.~\ref{fig:mu1km}
shows the distribution of the energy loss of muons in running
over 1 km in ice. The energy loss distribution due
to the pair creation may be narrow enough for the CEL approximation,
this is not the case for the distributions due to the other interaction,
however.
It is not appropriate to approximate
the entire distribution by a $\delta$-function, which implies
that the energy {\it loss} rather than the energy 
would be better to describe the event characteristics.

\begin{figure}
\includegraphics[width=.4\textwidth,clip=true]{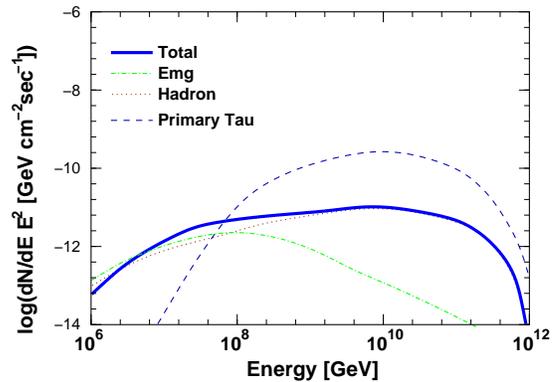}
\caption{\label{fig:gzk_4_4_casTau}
The tau fluxes at the IceCube depth originated
in the cosmogenic neutrinos as a function of energy (the dashed curve),
total energy loss in ice (the solid curve), energy loss in form
of electromagnetic cascades (the dash-dot curve), and that in
form of hadronic cascades (the dotted curve).}
\end{figure}

As more realistic criteria, we introduce the threshold
of the energy loss in ice instead of the energy itself.
Fig.~\ref{fig:atmMuon_casMu_angle_10PeV} shows
the GZK integral flux dependences on the zenith angle
in the case of 10PeV threshold of the energy loss. 
One can see in comparison to Fig.~\ref{fig:atmMuon_angle_10PeV} that
the GZK fluxes are larger than or comparable to
the muon background intensity in all the zenith directions
in this energy-loss-based criteria.
It indicates that it is probable that the EHE neutrino search
using downward events can be made under almost background-free
environment.

\begin{table}
\caption{\label{tab:rate} The event rates for several EHE neutrino models.
The notation of the model name is same as in Table~\ref{tab:event}.}
\begin{ruledtabular}
\begin{tabular}{l|ll}
   & $N_{\mu}(E_{loss}\geq 10PeV)$ & $N_{\tau}(E_{loss}\geq 10PeV)$\\
   & [km$^{-2}$ year$^{-1}$] & [km$^{-2}$ year$^{-1}$]\\\hline 
GZK Downward& 0.15 & 0.10 \\
GZK Upward & 0.0081 & 0.0083 \\
TD Downward & 1.18 & 0.93 \\
Atm. $\mu$ & 0.055 & - \\
Atm. $\mu$ (Corsika)& 0.016 & - \\
\end{tabular}
\end{ruledtabular}
\end{table}

It should be noted that the tau flux is lower than
the muon flux in this criteria. This is because the heavier
mass of tau suppresses the energy loss compared to that of 
muons with same energy. This situation is illustrated in
Fig.~\ref{fig:gzk_4_4_casTau} where the tau fluxes are
plotted as functions of energy and energy loss in ice
during 1km propagation. The intensity above $10^7$ GeV
is reduced because of the energy loss suppression.
The higher energy loss takes place in form of
the hadronic cascades initiated by the photonuclear
interaction. Table~\ref{tab:event} lists the intensity
of muon and tau above 10 PeV of energy loss for
the fluxes of the cosmogenic~\cite{yoshida93} and 
top-down model~\cite{SLBY}.
The event rate under this criteria 
is found to be 0.27 $(\mu+\tau)$/km$^2$ year for the cosmogenic neutrino
fluxes with the moderate source evolution. Note that
the downward event rate is 0.25 /km$^2$ year and dominates
in the overall rate. 
The event rates for the various neutrino production models
are summarized in Table~\ref{tab:rate}.

\begin{figure*}\
\includegraphics[width=.8\textwidth,clip=true]{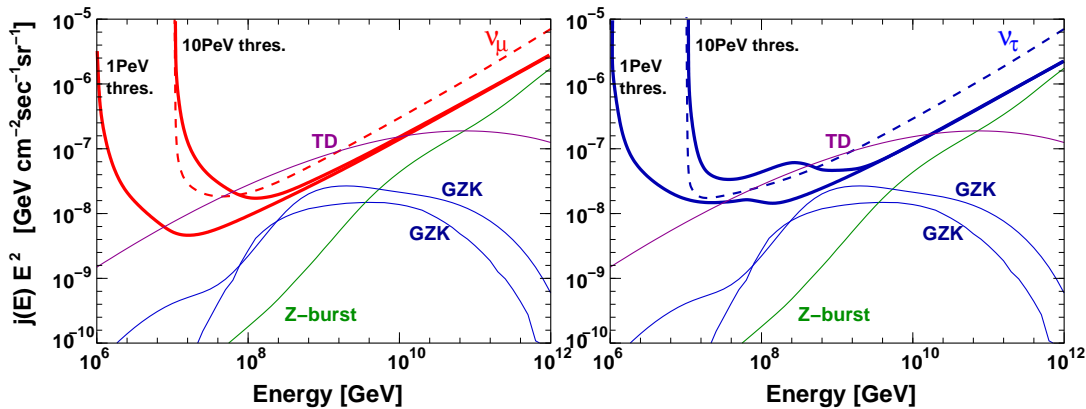}
\caption{\label{fig:iceCube_sensitivity}
The IceCube sensitivities on the EHE neutrino fluxes. 90 \% C.L.
limits by a 1km$^2$ detection area with 10 years observation 
are drawn. The left panel shows the case of $\nu_{\mu}$ and
the right panel shows the $\nu_{\tau}$ case. 
Labels refer to GZK (\cite{yoshida93} for the lower curve,
\cite{kalashev02} for the upper curve), TD~\cite{SLBY},
and  Z-burst~\cite{yoshida98}. The dashed curves show
the sensitivities by events of neutrinos interacting inside
the detector volume.}
\end{figure*}

The IceCube sensitivity on EHE neutrinos can be evaluated
by the event rate per energy decade $dN/dLogE$.
For a given energy of primary neutrinos, the secondary
muon and tau fluxes are calculated by the transport equations
Eqs.~\ref{eq:transport1} and~\ref{eq:transport2}
as a function of zenith angles.
The probability that energy loss with the threshold value
or greater occurs are estimated by the Monte Carlo simulation
and convoluted with the flux integration over energy and zenith
angle to give the rate. Fig.~\ref{fig:iceCube_sensitivity}
shows the resultant sensitivity by the IceCube detector
with 1 km$^2$ detection area. The various model predictions
are also shown for comparison. The 90 \% C.L. upper limit
{\it i.e.}, 2.3 event/energy decade/10 year is plotted
for 10 PeV and 1 PeV threshold of the energy loss, respectively.
The $\nu_{\mu}$ sensitivity is better than that of $\nu_{\tau}$
below $10^{8}$ GeV region because muon energy loss in a detection
volume is larger than
that of taus with same energy, but the tau decay which
results in large energy deposit in a detection volume
make dominant contributions in the $\nu_{\tau}$ sensitivity
in this relatively low energy range forming a slight
bump structure in the sensitivity curve.
The 90 \% C.L. upperlimit of
EHE neutrino fluxes by a km$^2$ detection area
would be placed at
 $E^2dF/dE\simeq 3.7\times 10^{-8}$ GeV/cm$^2$ sec sr for $\nu_{\mu}$ 
and $4.6\times 10^{-8}$ for $\nu_{\tau}$ with energies of $10^9$ GeV
in absence of signals with energy-loss in a detection volume of 10PeV or greater.

This bound would not exclude 
the cosmogenic neutrino production model but strongly
constrain the cosmic ray injection spectrum in the model.
Cosmic ray nucleon injection spectra harder than $E^{-1.5}$
would violate the bound~\cite{kalashev02}. On the other hand,
as long as the injection spectra is softer than $E^{-2}$,
which is very likely in case of the astrophysical cosmic ray sources,
the IceCube bound would less constrain the source evolution
as seen in Fig.~\ref{fig:iceCube_sensitivity} where we plotted
an extreme scenario of the cosmological evolution $(1+z)^5$ where
$z$ is the redshift~\cite{kalashev02}. 
Stronger evolution possibilities than
this case are inconsistent
with the diffuse background 
$\gamma$-ray observation by EGRET~\cite{sreekumar97}
since the GZK mechanism also 
initiated the electromagnetic cascades~\cite{yoshida93,proth96,kalashev02}
via photoproduced $\pi^0$ decay and $e^{\pm}$ pair creation by
EHECRs collisions with the CMB photons,
forming the photon flux below 100 GeV which is constrained
by the observation.

The topological defects scenario, on the other hand, would
be severely constrained by absence of EHE event detection
by the IceCube. The expected event rate is $\sim$ 2 events/year km$^2$
as one can calculate from Table~\ref{tab:event}.
The expected EHE neutrino flux
in the Z-burst model~\cite{fargionweiler97}, the scenario that
the collisions of EHE
neutrinos with the cosmological background
neutrinos to explain the EHECR fluxes without the GZK cutoff,
is well below the IceCube bound if the injection neutrino
spectrum is $E^{-1}$ as described in Ref.~\cite{yoshida98}.

Although less significant, there are $\mu$ and $\tau$ events
produced by neutrinos {\it inside} the detector instrumented  volume.
In this case the produced charged leptons propagate
only a part of the observation volume. We carried out the same
Monte Carlo simulation deriving the results of Fig.~\ref{fig:mu1km}
but in which $\nu_{\mu}$ and $\nu_{\tau}$ were initially entering into
the ice volume. The probability that neutrinos interact inside
the 1 km$^3$ volume and that the produced muon or tau losses its
energy greater than 10 PeV were estimated and convoluted
with the neutrino intensity at the IceCube depth.
The detection sensitivities by this channel
are shown in thick dashed curves in Fig.~\ref{fig:iceCube_sensitivity}.
In the EHE regime above $\sim 10^{8}$ GeV, the intensity of internally produced 
muon and tau events is too small to contribute the overall sensitivity
because the neutrino target volume is limited by the size of the detector
{\it i.e.} 1 km$^3$. Below $10^{8}$ GeV, on the other hand, 
including this channel improves
the sensitivity in sizable manner because the energy losses of muons and taus
during their propagation over long distances are more likely to transfer
them out of the energy range above the 10 PeV threshold,
which leads to reduction of the effective neutrino target volume
for producing EHE muons and taus outside the detector volume.
There is little gain in EHE neutrino searches, however, because
the proposed EHE neutrino models have its main energy range above
$10^{8}$ GeV.

\section{\label{sec:summary}Summary and Outlook}

We calculated propagation of 
the EHE neutrinos and charged leptons in the earth to
derive their intensities and their dependence on
nadir angles. The secondary produced muons and taus
form detectable fluxes at the IceCube depth, with
intensity of three order of magnitude lower than
the neutrino fluxes. The realistic criteria,
requiring energy deposit grater than 10 PeV in
1 km$^3$ volume of ice, leads to $\sim$ 0.27 events/year
for the cosmogenic neutrinos in case of the moderate
source evolutions. The topological defects scenario
would be severely constrained. 

The atmospheric muon background are likely to be
negligible even for downgoing events. The background rate
is $\sim 0.05$ event/km$^2$ year. It should be noted
that we ignored the possible contributions of
prompt muons from the charm decay
in EHE cosmic ray air showers~\cite{promptMu}, however. 
The atmospheric muon intensity can be 
increased by an order of magnitude
but a large uncertainty remains due to highly uncertain
cross sections in the charm production.
The mass composition of cosmic rays in the relevant energy range
would also be a deciding factor of the prompt muon flux intensity:
It can be reduced by an order of magnitude if the cosmic rays
are heavy nuclei~\cite{roulet03} and may not constitute a background
in the EHE neutrino search. Even if the prompt muon intensity
is sizable, their energy spectrum
would still be much steeper than the expected spectrum
in the proposed EHE cosmic neutrino models, however,
and one can easily distinguish the signal detections
from the prompt muon background events if the neutrino observatory
has reasonable resolution for the energy loss of muons and taus
tracks. The detector resolution issues require
the detailed detector Monte Carlo simulations
for further investigations. The AMANDA experience
in relatively low energy muon reconstruction would lead
to energy resolution of $\Delta$ Log E $\simeq 0.3$~\cite{IceCube_perform}.
The development of the detector Monte Carlo simulation
is under progress and its application to the present results
will be an important future work toward the search for
EHE neutrinos by the IceCube observatory.

\begin{acknowledgments}
We wish to acknowledge the IceCube collaboration
for useful discussions and suggestions.
We thank Dmitry Chirkin, Thomas Gaisser, and Esteban Roulet
for helpful discussions on the atmospheric muon issues.
We also thank John Beacom and Igor Sokalski for their
valuable comments.
Special thanks go to Mary Hall Reno for
providing the cross section table evaluated by
the CTEQ version 5 parton distribution.
This work was supported in part by
the Grants-in-Aid (Grant \# 15403004 and 15740135) in 
Scientific Research from the MEXT (Ministry of Education, 
Culture, Sports, Science, and Technology) in Japan.
\end{acknowledgments}




\begin{thebibliography}{99}

\bibitem{nagano01} for a review see, e.g.,
M.~Nagano and A.~A.~Watson, Rev. Mod. Phys. {\bf 72}, 689 (2000);
J.~W.~Cronin, Rev.~Mod.~Phys. {\bf 71}, S165 (1999);
S.~Yoshida and H.~Dai, J.~Phys.~G:~Nucl.~Part.~Phys. {\bf 24}, 905 (1998).

\bibitem{berezinsky69}
V.~S.~Beresinsky and G.~T.~Zatsepin, Phys.~Lett. {\bf 28B}, 423 (1969).

\bibitem{GZK} K.~Greisen, Phys.~Rev.~Lett.~{\bf 16}, 748 (1966);
G.~T.~Zatsepin and V.~A.~Kuzmin, Pisma Zh.~Eksp.~Teor.~Fiz.~{\bf 4}, 114
(1966) [JETP.~Lett.~{\bf 4}, 78 (1966)].

\bibitem{BHS} P.~Bhattacharjee, C.~T.~Hill, and D.~N.~Schramm,
Phys.~Rev.~Lett.~{\bf 69}, 567 (1992).

\bibitem{SLSC} G.~Sigl, S.~Lee, D.~N.~Schramm, and P.~S.~Coppi,
Phys. Lett. B {\bf 392}, 129 (1997).

\bibitem{SLBY} G.~Sigl, S.~Lee, P.~Bhattacharjee, and S.~Yoshida,
Phys.~Rev.~D {\bf 59}, 043504 (1999).

\bibitem{halzen01} J.~Alvarez-Mu\~{n}iz and F.~Halzen,
Phys.~Rev.~D {\bf 63}, 037302 (2001).

\bibitem{reno03} J.~Jones, I.~Mocioiu, M.~H.~Reno, and I.~Sarceiv,
hep-ph/0308042.

\bibitem{feng02} J.~L.~Feng, P.~Fisher, F.~Wilczek, and T.~M.~Yu,
Phys.~Rev.~Lett.~{\bf 88}, 161102 (2003);
K.~Giesel, J.~-H.~Jureit and E.~Reya,
Astropart.~Phys. {\bf 20}, 335 (2003).

\bibitem{beacom02}
J.~F.~Beacom, P.~Crotty and E.~W.~Kolb,
Phys.~Rev.~D {\bf 66}, 021302 (2002).

\bibitem{IceCube} S.~Yoshida, Proc. of the 28th ICRC, H.E.2.3, 1369 (2003);
http://icecube.wisc.edu/.

\bibitem{kelner00} S.~Kelner, R.~P.~Kokoulin, A.~A.~Petrukhin,
Phys. At. Nucl. {\bf 63}, 1603 (2000).

\bibitem{PC} R.~P.~Kokoulin and A.~A.~Petrukhin, 
Proc. of the 12th ICRC (Hobert) Vol.6 p.A2436 (1971);
S.~R.~Kelner, R.~P.~Kokoulin, and A.~A.~Petrukhin, 
Phys. At. Nucl. {\bf 62}, 1894 (1999).

\bibitem{BS} Yu.~M.~Andreev, L.~B.~Bezrukov, and E.~V.~Bugaev,
Phys. At. Nucl. {\bf 57}, 2066 (1994);
S.~R.~Kelner, R.~P.~Kokoulin, and A.~A.~Petrukhin, 
Phys. At. Nucl. {\bf 60}, 576 (1997);
I.~A.~Sokalski, E.~V.~Bugaev, and S.~I.~Klimushin,
Phys.~Rev.~D {\bf 64}, 074015 (2001).

\bibitem{PN} 
S.~IyerDutta, M.~H.~Reno, I.~Sarcevic, and D.~Seckel, 
Phys.~Rev.~D {\bf 63}, 094020 (2001).

\bibitem{ALLM}
H.~Abramowicz and A.~Levy, hep-ph/9712415.

\bibitem{gandhi96} R. Gandhi, C. Quigg, M. H. Reno, and I. Sarcevic,
Astropart.\ Phys.\ {\bf 5}, 81 (1996); Phys.~Rev.~D {\bf 58} 093009 (1998).


\bibitem{cteq} CTEQ Collaboration, H. Lai {\it et al.},
Phys.~Rev.~D {\bf 55}, 1280 (1997).

\bibitem{gaisser90}
T.~K.~Gaisser, {\it Cosmic Rays and Particle Physics},
(Cambridge University Press, Cambridge, England, 1990).

\bibitem{TAU}
S.~I.~Dutta, M.~H.~Reno, I.~Sarcevic,
Phys.~Rev.~D {\bf 62}, 123001 (2000).

\bibitem{PEM} A. M. Dziewonsky and D. L. Anderson,
Phys.~Earth.~Planet.~Inter {\bf 25}, 297 (1981);
S.~V.~Panasyuk, http://cfauvcs5.harvard.edu/lana/rem/index.html.

\bibitem{yoshida93}
S.~Yoshida and M.~Teshima, Prog.~Theor.~Phys. {\bf 89}, 833 (1993).

\bibitem{proth96}
R.~J.~Protheroe and P.~A.~Johnson, Astropart.~Phys. {\bf 4}, 253 (1996).

\bibitem{seckel01} R.~Engel, D.~Seckel, T.~Stanev,
Phys.~Rev.~D {\bf 64}, 093010 (2001).

\bibitem{kalashev02} O.~E.~Kalashev, V.~A.~Kuzmin, D.~V.~Semikoz,
and G.~Sigl, Phys.~Rev.~D {\bf 66}, 063004 (2002).

\bibitem{SK00} SuperKamiokande Collaboration, Y.~Fukuda {\it et al.},
Phys.~Rev.~lett. {\bf 85}, 3999 (2000).

\bibitem{igor2003} I.~Sokalski, private communications (2003).

\bibitem{ATM96} V.~Agrawal, T.~K.~Gaisser, P.~Lipari, and T.~Stanev,
Phys.~Rev.~D {\bf 53}, 1314 (1996).

\bibitem{Corsika} D.~Heck {\it et al.}, 
Report FZKA {\bf 6019}, (Forschungszentrum Karlsruhe 1998).

\bibitem{sreekumar97} P. Sreekumar {\it et al.},
Astrophys. J. {\bf 494}, 523 (1998).

\bibitem{fargionweiler97} D. Fargion, B. Mele, and A. Salis,
Astrophys.~J. {\bf 517}, 725 (1999); 
T.~J.~Weiler, Astropart.~Phys. {\bf 11}, 303 (1999).

\bibitem{yoshida98} 
S. Yoshida, G. Sigl, and S. Lee, Phys.\ Rev.\ Lett. {\bf 81}, 5505 (1998).

\bibitem{promptMu}
See e.g., T.~S.~Sinegovskaya and S.~I.~Sinegovsky,
Phys.~Rev.~D {\bf 63}, 096004 (2001), and references therein.

\bibitem{roulet03}
J.~Candia and E.~Roulet, astro-ph/0306632.

\bibitem{IceCube_perform} 
IceCube Collaboration, J.~Ahrens {\it et al.}, astro-ph/0305196,
accepted for publication in Astropart.~Phys.
\end{thebibliography}
\end{document}